\documentclass[twocolumn,usenatbib,prl,floats,floatfix,usegraphicx]{revtex4}

\usepackage{epsfig}
\usepackage{graphicx}
\usepackage{graphics}

\usepackage{amsfonts}  % For font for constants in math.
\usepackage{amsmath, amssymb, stmaryrd}     %for math symbols

\usepackage{psfrag}  % To change labels in plots.

\usepackage{pict2e}

\def\be{\begin{equation}}
\def\ee{\end{equation}}
\def\bi{\begin{itemize}}
\def\ei{\end{itemize}}

\begin{document}
\title{Cosmological simulations of screened modified gravity out of the static approximation: effects on matter distribution}
\author{Claudio Llinares}
\affiliation{Institute of theoretical astrophysics, University of Oslo, 0315 Oslo, Norway}
\author{David F. Mota}
\affiliation{Institute of theoretical astrophysics, University of Oslo, 0315 Oslo, Norway}

\begin{abstract}
In the context of scalar tensor theories for gravity, there is a universally adopted hypothesis when running N-body simulations that time derivatives in the equation of motion for the scalar field are negligible.  In this work we propose to test this assumption for one specific scalar-tensor model with a gravity screening mechanism: the symmetron.  To this end, we implemented the necessary modifications to include the non-static terms in the N-body code Ramses.  We present test cases and results from cosmological simulations.  Our main finding when comparing static vs. non-static simulations is that the global power spectrum is only slightly modified when taking into account the inclusion of non-static terms.   On the contrary, we find that the calculation of the local power spectrum gives different measurements. Such results imply one must be careful when assuming the quasi-static approximation when investigating the environmental effects of modified gravity and screening mechanisms in structure formation of halos and voids  distributions.

\end{abstract}
%galaxies: kinematics and dynamics- methods: N-body simulations}

\maketitle
\section{Introduction}
General Relativity (GR) can be considered as the foundation stone of the standard model for cosmology ($\Lambda$CDM).  Indeed, the assumption of this theory as valid leads to the need of the two building blocks of the model: dark matter and dark energy.  The model is able to match large numbers of observables on large scales.  However, the nature of the two dark components is still unclear.  Among the different solutions to the philosophical and quantitative issues associated to this components, exists the idea of modifying the gravitational theory \citep[][]{2012PhR...513....1C}.  As GR was proven to be valid in solar system scales, any modification introduced must fullfil the requirement of reducing to GR in these scales, which is done through screening mechanisms.  Within the context of scalar-tensor theories, there are three of such mechanisms based on conformal couplings (Vainshtein \citep{vain}, Symmetron \citep[][]{{2010PhRvL.104w1301H}} and Chameleon \citep[][]{{cham}}).  In addition to this, \citet[][]{2012PhRvL.109x1102K} recently proposed a mechanism which is based on a disformal coupling.

The search for predictions for this kind of theories in the scales of galaxies and clusters of galaxies (i.e. in the non-linear regime of cosmological evolution) requires the use of cosmological simulations.  Several works exist in the literature in the context of scalar tensor theories for gravity \citep[e.g.][]{{2008PhRvD..78l3523O}, schmidt_dgp_code, baldi_coupled_quintessence_code,lmb1,lmb2,lmb3, zhao_baojiu_forf_code, {2012JCAP...01..051L}, baldi_de_sim, {2012MNRAS.422.1028B}, 2012JCAP...10..002B, 2013MNRAS.436..348P, nbody_chameleon, dgp_code_durham,li1,li2,2014A&A...562A..78L, 2010PhRvD..82j3536H, 2013MNRAS.435.2806H, 2013MNRAS.431..749C, 2013arXiv1310.6986C}.  The main assumption in this papers is that the quasi-static limit is a good approximation and thus, time derivatives can be neglected in the equation of motion of the scalar field.  Within the context of standard gravity, it can be shown analytically that the static equations are valid even outside the horizont and thus, Newtonian simulations do not give just a good approximation, but the right solution at all scales \citep[][]{2013arXiv1308.0057R, 2011PhRvD..83l3505C, 2012PhRvD..85f3512G}.  On the other side, the validity of the static limit for the scalar field equation is still unclear.  
%While in the approximation is known to be valid for GR even at scales comparables with the horizon \citep[][]{2013arXiv1308.0057R, 2011PhRvD..83l3505C, 2012PhRvD..85f3512G}, the validity for the scalar field equation is still unclear.  
The only work in this subject \citep[][]{2013arXiv1310.3266N} applies only to linear evolution.  In the non-linear case, first N-body simulations including time derivatives in the equation of motion of the scalar field were presented in Ref. \citep[][]{2013PhRvL.110p1101L} in the context of the symmetron model.  However, in there, only the properties of the solutions for the scalar field were studied and there was no mention of the impact that these new solutions have in the matter distribution.  In other words, the simulations were run without including the effects of the fifth force associated to the scalar field in the geodesics equation.  The question wether observables such as the power spectrum of density perturbations are affected by the non-static terms in the non-linear regime is still open.  The aim of this paper is to test the existence of these effects.  To this end, we included to non-static terms of the Klein-Gordon equation in the code Isis \citep[][]{2014A&A...562A..78L}, which is modification of the code Ramses \citep[][]{2002A&A...385..337T} that includes scalar fields in its static limit.  We focus this paper in the symmetron screening mechanism \citep[][]{2010PhRvL.104w1301H}.  However our techniques can be easily generalized to different models.  In particular, there is an interesting family of models such as disformal gravity \citep[][]{2012PhRvL.109x1102K}, in which the screening of the fifth forces is directly related to the time derivatives of the scalar field and thus, can not be simulated assuming the static approximation.  

The paper is organized as follows:  section 2 describe the set of equations used in the N-body code, the method that we employ to solve them and details of the implementation in the code Ramses.  The tests that were made to this implementation are shown in section 3.   Section 4 and 5 describe 3D cosmological simulations that we run including the non-static scalar field as well as results that were obtained from them on the power spectrum of the density perturbations.  Last section includes our conclusions and discussion.

\section{The equations and the method}

\subsection{The symmetron equations}

The symmetron model is defined by the following action:
\begin{align}
\nonumber
S  = \int & \sqrt{-g} \left[ \frac{M^2_\mathrm{pl}}{2} R - \frac{1}{2}\nabla^a\phi \nabla_a \phi - V(\phi)\right] d^4x + \\
& + \int L_M(\tilde{g}_{\mu\nu}) d^4x, 
\label{action_scalar_tensor}
\end{align}
where the Einstein and Jordan frames metrics ($g_{\mu\nu}$ and $\tilde{g}_{\mu\nu}$) are related according to
\be
\tilde{g}_{\mu\nu} = A^2(\phi) g_{\mu\nu}.
\ee
The potential $V$ and conformal factor $A$ that define this particular model are:
\begin{align}
V(\phi) &= -\frac{1}{2}\mu^2\phi^2 + \frac{1}{4}\lambda\phi^4 \\
A(\phi) &= 1 + \frac{1}{2}\left(\frac{\phi}{M}\right)^2, 
\end{align}
where $\mu$ and $M$ are mass scales and $\lambda$ is a dimensionless constant.  The equation of motion for the scalar field that result from this action is:
\be
\nabla^a\nabla_a\phi = V_{,\phi} - A_{,\phi} A^3 \tilde{T},
\label{eq_motion_covariant}
\ee
where $\tilde{T}$ is the trace of the Jordan frame energy momentum tensor, which is defined as $\tilde{T}_{ab}=-(2/\sqrt{-\tilde{g}})\delta L_M/\delta\tilde{g}^{ab}$.  In order to introduce eq.\ref{eq_motion_covariant} in the N-body code we need to specify the metric, which we choose as a flat Friedmann-Lema$\hat{\text{i}}$tre-Robertson-Walker metric with only scalar perturbations:
\be
ds^2 = -(1+2\Phi)dt^2 + a^2(t) (1-2\Phi)(dx^2+dy^2+dz^2). 
\label{metric}
\ee
 With this metric, the equation of motion for the scalar field takes the following form
\be
\ddot{\phi} + 3 H \dot{\phi} - \frac{1}{a^2}\nabla^2\phi = -V_{,\phi} - A_{,\phi} \rho = -V_{eff,\phi}, 
\label{canonical_equation}
\ee
where we assumed that the conformal factor $A$ is close to one.  The dots in previous expression represent derivatives with respect to cosmic time $t$, $\rho$ is the matter density and the effective potential takes the following form:
\be
V_{s,eff}(\phi) = \frac{1}{2} \left(\frac{\rho}{M^2} - \mu^2 \right) \phi^2 + \frac{1}{4}\lambda \phi^4.
\label{effective_potential}
\ee
At this point, it is convenient to define a characteristic density:
\be
\rho_{SSB} = M^2 \mu^2. 
\ee
For values of the local density smaller than $\rho_{SSB}$, the scalar field is free to oscillate around a minimum which is different from zero and thus, a fifth force arises.  For densities that are larger than this value, the symmetry $\phi\rightarrow -\phi$ is restored and the scalar field oscillates around zero or, in other words, it is screened.  
For numerical convenience, we normalize the field $\phi$ with the minimum of the potential $\phi_{0}$ that corresponds to zero density and is given by:
\be
\phi_{0}^2 = \frac{\mu^2}{\lambda}.
\ee
By dividing eq.\ref{canonical_equation} by $\phi_{0}$ and defining the dimensionless quantity
\be
\chi=\frac{\phi}{\phi_{0}}, 
\ee
we obtain the equation of motion written in the following form:
\be
\label{eq_motion_chi}
\ddot{\chi} + 3H\dot{\chi} - c^2\frac{\nabla^2\chi}{a^2} = -\frac{c^2}{2\lambda_0^2} \left[ \frac{a_{SSB}^3}{a^3}\chi\eta - \chi + \chi^3  \right], 
\ee
where $\eta$ is the matter density field normalized with the background density, and 
\be
\lambda_{0} = \frac{1}{\sqrt{2} \mu}
\ee
 is the range for the field that corresponds to zero density.  We also defined a redshift of symmetry breaking $z_{SSB}$ (or it associated expansion factor $a_{SSB}$), with corresponds to the redshift at which the mean density of the universe is equal to the density of symmetry breaking $\rho_{SSB}$.

The code Ramses uses supercomoving coordinates \citep[][]{1998MNRAS.297..467M}, which are defined by:
\begin{align}
d \tau & = \frac{1}{a^2} dt\\
\tilde{\Phi} & = a^2 \Phi. 
\end{align}
To this change, we also add the following definition:
\be
\tilde{\chi} = a \chi
\label{change_chi_a_chi}
\ee
which will help in removing an explicit dependence with $a$ in the term related to the fifth force in the geodesics equation.  In this coordinates, the equation of motion for the scalar field takes the following form:
\be
\tilde{\chi}'' - \tilde{H}\tilde{\chi}' - \tilde{H}'\tilde{\chi} - a^2 c^2 \nabla^2\tilde\chi = -\frac{a^4c^2}{2\lambda_0^2} \left[ \frac{a_{SSB}^3}{a^3}\tilde{\chi}\eta - \tilde{\chi} + \frac{\tilde{\chi}^3}{a^2}  \right],
\label{eq_symm_supercomoving}
\ee
where the prime denotes derivatives with respect to the supercomoving time $\tau$ and $\tilde{H} = a'/a$ is the supercomoving Hubble factor.

The metric perturbations $\Phi$ will be solution of the following equation:
\be
\nabla^2 \Phi = \frac{3}{2}\frac{\Omega_m H_0^2}{a} \delta, 
\ee
where $\delta$ is the over-density defined as $\delta\rho/\rho_b$, $\rho_b$ is the background density, $\Omega_m$ is the background density in terms of the critical density of the universe and $H_0$ is the Hubble constant.

In this paper, we will track only the dark matter component, which can be described by means of free particles whose coordinates follow the geodesics equation.  After normalizing the field and defining $\beta=\phi_0 M_{\text{pl}}/M^2$, we obtain the geodesics written in the following form:
\begin{multline}
\frac{d^2\mathbf{x}}{d\tau^2} + \nabla\tilde{\Phi} + \\
6\Omega_m H_0^2 \frac{(\beta\lambda_0)^2}{a_{SSB}^3}\left[ \tilde{\chi}\nabla\tilde{\chi} +\frac{1}{c^2 a^3} \tilde{\chi}\left(\tilde{\chi}'-\tilde{H}\tilde{\chi}\right)\mathbf{x}'\right]= 0.
\label{symmetron_geodesics_tilde_chi}
\end{multline}
Note that the second term inside the squared bracket is lower order with respect to the first one if one think in an expansion in terms of the speed of light $c$.  Thus, during the rest of this work we will neglect the term and include the effects of the fifth force only through the term $\tilde{\chi}\nabla\tilde{\chi}$.  The study of the effects of high order terms is left for future work.

\subsection{Solving the non-static field equation}

The standard way to solve eq.\ref{eq_motion_chi} in the context of cosmological simulations is to assume that the terms responsible for the oscillations of the scalar field are small and thus its solution can be approximated by the solution of:
\be
\label{static_eq_motion_chi}
\frac{\nabla^2\chi}{a^2} = \frac{1}{2\lambda_0^2} \left[ \frac{a_{SSB}^3}{a^3}\chi\eta - \chi + \chi^3  \right].
\ee
which can be obtained, for instance, by using multigrid methods.  In this paper we go beyond this approximation and solve the complete equation, including the time derivatives of the scalar field.  The method applied to cosmological simulations was presented in \citet[][]{2013PhRvL.110p1101L} and exploits the fact that the Klein-Gordon equation is formally equivalent to the geodesics equation.  Thus, one can apply a leap-frog scheme, not to positions and velocities of a set of particles, but to the scalar field $\tilde{\chi}$ and its time derivative.  The definition
\be
q = \frac{\tilde{\chi}'}{a}
\ee
gives the following set of first order equations:
\begin{align}
\label{eq_1_of_system}
q' & = \frac{\tilde{H}'}{a} \tilde{\chi} + a c^2\nabla^2\tilde{\chi} - \frac{a^3c^2}{2\lambda_0^2}\left[\frac{a_{SSB}^3}{a^3}\eta\tilde{\chi} - \tilde{\chi} + \frac{\tilde{\chi}^3}{a^2}  \right] \\
\label{eq_2_of_system}
\tilde{\chi}' & = a q.
\end{align}
By using second order discretization in time and implementing a leap-frog scheme, we obtain the following evolution equations for the time step $n$:
\begin{align}
\tilde{q}_{n+1/2} & = \tilde{q}_n + \\
\nonumber 
& \left\{c^2 a_n \nabla^2\tilde{\chi}_n - \right.\\
\nonumber
& \left.\frac{c^2a_n^3}{2\lambda_0^2} \left[ \frac{a_{SSB}^3}{a_n^3}\tilde{\chi}_n\eta_n - \tilde{\chi}_n + \frac{\tilde{\chi}_n^3}{a_n^2}  \right] + \frac{\tilde{H}'_n}{a_n}\tilde{\chi}_n \right\} \frac{\Delta \tau}{2}\\
\tilde{\chi}_{n+1} & = \tilde{\chi}_n + \left[ a_{n+1/2} \tilde{q}_{n+1/2}\right] \Delta \tau \\
\tilde{q}_{n+1} & = \tilde{q}_{n+1/2} + \\
\nonumber 
& \left\{c^2 a_{n+1} \nabla^2\tilde{\chi}_{n+1} - \right.\\
\nonumber
& \left. \frac{c^2a_{n+1}^3}{2\lambda_0^2} \left[ \frac{a_{SSB}^3}{a_{n+1}^3}\tilde{\chi}_{n+1}\eta_{n+1} - \tilde{\chi}_{n+1} + \frac{\tilde{\chi}_{n+1}^3}{a_{n+1}^2}  \right] + \right.\\
\nonumber
& \left. \frac{\tilde{H}'_{n+1}}{a_{n+1}}\tilde{\chi}_{n+1} \right\} \frac{\Delta \tau}{2}, 
\end{align}
where we divided the evolution of the variable $q$ in two small time steps as it is done in the standard Ramses code for the velocities of the particles.  See Ref.\citep[][]{llinares_thesis} for the application of the same scheme to the solution of the growth equation of linear density perturbations in the modified gravity case.  A similar approach was also applied in the context of scalar fields that are not coupled to matter \citep[][]{1989PhRvD..40.1002W}.

Initial conditions have to be determine for the scalar field and its time derivative.  The scalar field is expected to be screened in the early universe and thus, we chose $\tilde{\chi}_\mathrm{initial}=0$ and $q_\mathrm{initial}=n$, where $n$ is a small uniformly distributed random number.

During cosmological evolution the scalar field oscillates with a period that is much shorter than the time scales associated to the evolution of matter and the metric itself. In other words, a cosmological simulation run with a non-static scalar field adds one more time scale to the problem.  In order to avoid re-calculating gravitational forces more than is needed, we included a new time step which is used to advance only the scalar field within each of the standard time steps.  We determine this new time step by  estimating the period of the oscillations that are associated to a uniform density field given by the maximum value in the box.  Under this conditions and neglecting the expansion term, the equation of motion of the scalar field (eq.\ref{eq_symm_supercomoving}) takes the following form:
\be
\tilde{\chi}'' = -\frac{a^4 c^2}{2\lambda_0^2} \left[\frac{a_{SSB}^3}{a^3}\eta-1 \right]\tilde{\chi}, 
\ee
where we assume that the density is above symmetry breaking and that the oscillations are small, which means that one can approximate the effective potential with a second order polinomial.  Proposing a solution with the following form:
\be
\tilde{\chi} = A \exp(i\omega\tau)
\ee
and defining the period as:
\be
P = \frac{2\pi}{\omega}, 
\ee
we obtain:
\be
\label{period_symm}
P = 2\pi \frac{\sqrt{2}\lambda_0}{a^2 c} \frac{1}{\sqrt{\frac{a_{SSB}^3}{a^3}\eta-1}}.
\ee
The time step for the scalar field is then defined as a given fraction of this period $P$, which was calibrated during the testing phase of the development of the code.

For details in the implementation, we follow the algorithm included in the standard Ramses code \citep[][]{2002A&A...385..337T}.  The following pseudo-code describes in detail the complete algorithm for a given time step $N$:
\begin{tabbing}
\hskip\parindent\=\hskip\parindent\=\hskip\parindent\=\hskip\parindent\=\kill
%$N=n=0$ \\
%$\Delta\mathrm{T} = \Delta\tau = 0$ \\
%while(1) \\
%\{  \+ \\
Calculate $\delta_{N} = f(x_{N}) $ \\
Solve Poisson's equation for $ \tilde{\Phi}_{N} = f( \delta_{N},  a_{N} )$ \\
// \textit{Do second half of previous step} \\
$p_{N} = f(p_{N-1/2}, \nabla\tilde{\Phi}_{N}, \tilde{\chi}_{n}, \nabla\tilde{\chi}_n, \Delta\mathrm{T}/2)$\\
// \textit{Do step $N$} \\
$\mathrm{Determine} ~ \Delta\mathrm{T}$ \\
$\mathrm{T} = \mathrm{T} + \Delta\mathrm{T} \Rightarrow \mathrm{Determine} ~ a_{N+1}$ \\
$p_{N+1/2} = f(p_N, \nabla\tilde{\Phi}_N, \tilde{\chi}_{n}, \nabla\tilde{\chi}_n, \Delta\mathrm{T}/2)$ \\
$x_{N+1} = f(x_N, p_{N+1/2}, \Delta\mathrm{T})$  \\
call advance\_chi() 
%\- \\
%\}
\end{tabbing}

\begin{tabbing}
\hskip\parindent\=\hskip\parindent\=\hskip\parindent\=\hskip\parindent\=\kill
subroutine advance\_chi() \\
\{  \+ \\
$\Delta\tau_{old} = \Delta\tau$ \\
$\mathrm{Determine} ~ \Delta\tau = f(\delta_{N,\mathrm{max}}, a_N)$ and $A = \Delta T / \Delta \tau$ \\
%$\mathrm{Determine} ~ P = \Delta T / \Delta \tau$ and adjust $\Delta\tau$ \\
for $n$=1 to $A$ \\
\{ \+ \\
Calculate $\nabla^2\tilde{\chi}_{n}$ \\
// \textit{Do second half of previous step} \\
if($n==1$) \+ \\
$\tilde{q}_{n}  = f(\delta_N, \tilde{\chi}_{n}, \tilde{q}_{n-1/2}, \nabla^2\tilde{\chi}_{n}, a_{n}, \Delta\tau_{old}/2)$ \- \\
else \+ \\
$\tilde{q}_{n}  = f(\delta_N, \tilde{\chi}_{n}, \tilde{q}_{n-1/2}, \nabla^2\tilde{\chi}_{n}, a_{n}, \Delta\tau/2)$ \- \\
// \textit{Do step $n$} \\
$\tilde{q}_{n+1/2}  = f(\delta_N, \tilde{\chi}_n, \tilde{q}_n, \nabla^2\tilde{\chi}_n, a_n, \Delta\tau/2)$ \\
$\tau = \tau+\Delta\tau/2 \Rightarrow \mathrm{determine } ~ a_{n+1/2}$ \\
$\tilde{\chi}_{n+1} = f(\tilde{\chi}_n, \tilde{q}_{n+1/2}, a_{n+1/2}, \Delta\tau)$ \\
$\tau = \tau+\Delta\tau/2 \Rightarrow \mathrm{determine } ~ a_{n+1}$ \- \\
%$n=n+1$ \-\\
\} \-\\
\}
\end{tabbing}
Here, the symbol $f$ does not refer to a specific function, but only denotes functional dependence between the different variables.  The variable $T$ is the time that corresponds to the large time steps (i.e. those that correspond to the standard leap-frog included in the original Ramses code).  Note that the particles' position are updated using the instantaneous value of the scalar field.  A different approach could consist in using the mean value of the scalar field taken over the large steps $N$ or to kick the particles in every short time step.  The impact of different approaches in the final solution is beyond the scope of this paper and left for future work. 

\section{Tests}

In order measure the quality of our numerical solutions of the full equation of motion of the scalar field, we compare them with solutions obtained using a Runge-Kutta algorithm in two different contexts:  with and without linearizing of the equations.  It is important to mention that the change from a non-linear to a linear code involves only to change the source of equation \ref{eq_1_of_system} in the evolution scheme and thus, it is trivial to implement.  In this section we provide a description of a set of analytic solutions and compare them with the solutions provided by our new code.

\subsection{Uniform density}

The most straightforward test consist in studying the oscillations of the scalar field when the density field is uniform and equal to the mean density of the universe.  Analytic solutions can be obtained by linearizing the equation, however, for this test we prefer to keep the equation as it is in its original form (eq.\ref{eq_motion_chi}) and use a Runge-Kutta algorithm to obtain a solution that can be used to compare with.  The Runge-Kutta integration was made using the eight order solver with variable steps that is included in the open source library GSL \citep[][]{0954161734} and using the expansion factor as time variable.  During the test, the density $\eta$ was kept constant in time and equal to one.

\begin{figure}%[!h]
  \begin{center}
    \includegraphics[width=0.47\textwidth]{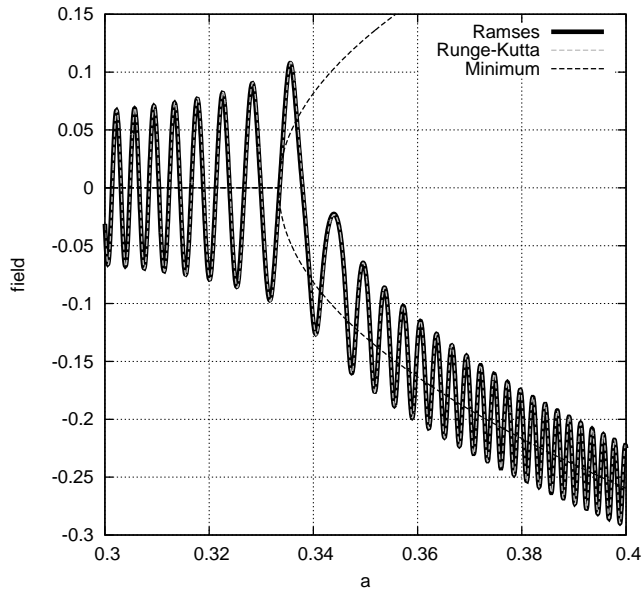}
    \caption{Result of the test with uniform density.  The continuous black and dashed gray lines correspond to the solution obtained with the new solver and the Runge-Kutta solution used for comparison.  The dashed black line shows the minimum of the effective potential.}
    \label{fig:test_constant_density}
  \end{center}
\end{figure}

The initial redshift of the test that we present here is $z=3.34$ and the initial value for the perturbed scalar field and its time derivative are $\tilde{\chi}=0.05$ and $q=0$.  Fig.\ref{fig:test_constant_density} shows the result of the test for the values that are close to $a_{SSB}$, when the scalar field changes from being screened to not screened.  The continuous black line is the result obtained with the modified 3D Ramses code, while the dashed gray line is the Runge-Kutta solution.  The dashed black line corresponds to the minimum of the effective potential which is given by:
\be
\tilde{\chi}_{min} = a \sqrt{1-\frac{a_{SSB}^3}{a^3}}.
\ee
Both solutions agree with each other, showing that the code can recover the oscillations of the background correctly.  

\subsection{Non-uniform density}

In order to test if the term that involves the Laplacian in eq.\ref{eq_1_of_system} is properly implemented, we repeated the test using a 1D non-uniform density distribution.  Runge-Kutta solutions to be used for comparison can be obtained in Fourier space after linearizing the equations.  The linearization can be made by assuming that:
\be
\frac{a_{SSB}^3}{a^3}\eta \ll 1,
\ee
which is valid in situations when the scalar field is not screened.  In this case, the scalar field $\tilde{\chi}$ will oscillate around its vacuum value which is equal to $a$ in the supercomoving frame.  Thus, the solution can be written as:
\be
\tilde{\chi} = a + \epsilon, 
\ee
where the perturbation $\epsilon$ is assumed to be much smaller than one.  By substituting this definition in eq.\ref{eq_symm_supercomoving} and neglecting high order terms we get:
\be
\epsilon'' - \tilde{H}\epsilon' - \tilde{H}'\epsilon - c^2a^2\nabla^2\epsilon = \frac{a^4c^2}{2\lambda_0^2}\left[\frac{a_{SSB}^3}{a^3}\eta + 2\epsilon \right].
\label{linear_supercomoving}
\ee
The change:
\be
u = \frac{\epsilon'}{a}
\ee
gives:
\begin{align}
\frac{d\epsilon}{d\tau} & = a q  \\
\frac{dq}{d\tau} & = \frac{H'}{a} \epsilon + ac^2\nabla^2\epsilon - \frac{a^3c^2}{2\lambda_0^2}\left[\frac{a_{SSB}^3}{a^3}\eta + 2\epsilon \right], 
\end{align}
which is the equation that we included in Ramses for the test.  Note that the difference between the linear equation (eq.\ref{linear_supercomoving}) and the equations \ref{eq_symm_supercomoving} used in the non-linear version is only in the source, and thus the algorithm used to solve them is the same in both cases.

The Fourier space version of eq.\ref{linear_supercomoving} is:
\be
 \hat{\epsilon}'' - \tilde{H}\hat{\epsilon}' - \tilde{H}'\hat{\epsilon} + c^2a^2k^2\hat{\epsilon} = \frac{a^4c^2}{2\lambda_0^2}\left[\frac{a_{SSB}^3}{a^3}\hat{\eta} + 2\hat{\epsilon} \right].
\ee
By defining:
\be
v = \frac{\hat{\epsilon}'}{a}, 
\ee
we can write:
\begin{align}
\frac{d\hat{\epsilon}}{d\tau} & = a v  \\
\frac{dq}{d\tau} & = \frac{H'}{a} \hat{\epsilon} - ac^2k^2\hat{\epsilon} - \frac{a^3c^2}{2\lambda_0^2}\left[\frac{a_{SSB}^3}{a^3}\hat{\eta} + 2\hat{\epsilon} \right], 
\end{align}
which is the system that we solve using the Runge-Kutta method.  The comparison of this solution with the one provided by our numerical code was made in real space.  To this end, we converted back this solution by using the open source library FFTW \citep[][]{FFTW05}.

\begin{figure*}
\includegraphics[width=1.0\textwidth]{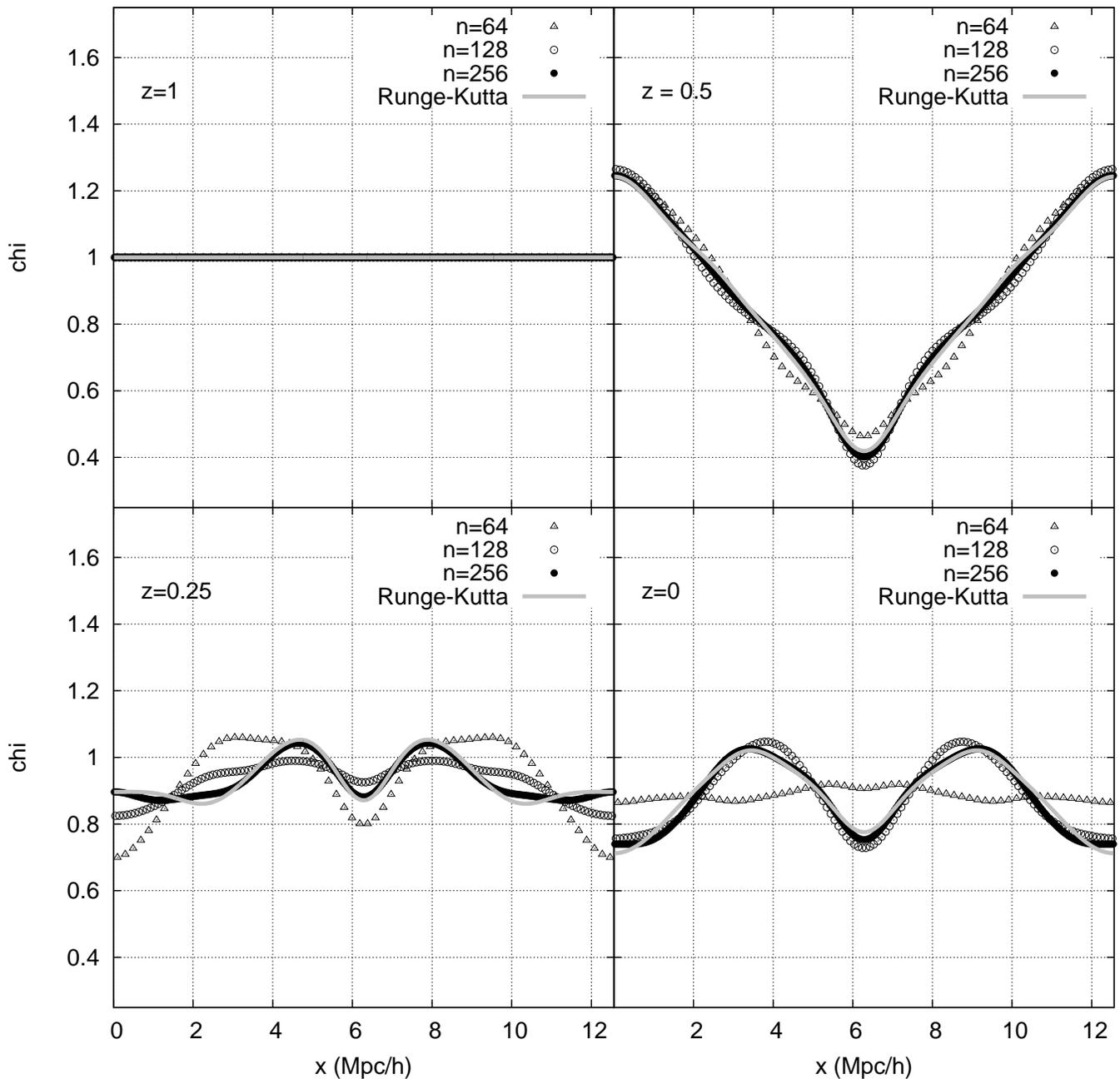}
\caption{Solution of the linear equation for a Gaussian density distribution at different stages.  The upper-left panel is the initial condition and the bottom-right the final result at redshift $z=0$.  The continuous line is the analytic solution and the points the numerical solution obtained using the 3D code with different resolutions.  The redshift that corresponds to every panel is shown in the upper left corner.  Here $z_{SSB}=1$.}
\label{fig:linear_sol_gauss}
\end{figure*}

\subsubsection{Choosing a density}

In order to fully specify the test to be made, we still need to fix a density distribution.  We use a Gaussian distribution with its maximum located the center of the box:
\be
\eta = A \exp\left(-(x-x_0)^2/b^2 \right).
\ee
To avoid dealing with translations in Fourier space, we obtained the Runge-Kutta solution by fixing $x_0=0$.  The constant $b$ was fixed to 1 Mpc and the normalization $A$ was specified by requiring the mean value of the overdensity to be equal to zero, which is equivalent to:
\be
<\eta> = \frac{1}{B^3} \int_{\text{Box}} \eta d^3x = 1,
\ee
where $B$ is the size of the box at redshift $z=0$.  This give us:
\be
A = \frac{B}{b\sqrt{\pi} \text{erf}\left(\frac{B}{2b}\right) },
\ee
where the erf function is defined as:
\be
\text{erf}(x) = \frac{2}{\sqrt{\pi}} \int_0^t \exp(-t^2) dt.
\ee

\subsubsection{Results of the test}

We run simulations with the above mentioned set up using the 3D code.  The density was defined on the grid using its analytic expression.  In order to keep its value constant, we decoupled during the test the scalar field solver from the time evolution of the particles.  We run three simulations with different resolutions with 64, 128 and 256 nodes per dimension.  To complete the set up, we need to specify the size of the time steps.  We use 60 steps in every period of the oscillations defined as in eq.\ref{period_symm}.  The simulations were run starting from redshift $z=z_{SSB}=1$ up to redshift $z=0$.  

Fig.\ref{fig:linear_sol_gauss} shows the result of the test at four different stages for the three different resolutions.  The upper-left and bottom-right panels show initial condition and final configuration at redshift $z=0$.  The continuous line is the Runge-Kutta solution and the points the solution extracted from the 3D box of the full simulation.  The overall form of the numerical solution agrees very well with the analytics.  Comparison between the three resolutions show that increasing resolution bring numerical and analytic solutions closer, which shows that the code converges to the right solution when increasing the resolution.  Furthermore, the plot shows that the low resolution runs, while can not reproduce exact details of the solution, give the same overall shape and mean value.

\section{Cosmological simulations}

In order to determine the consequences of including non-static terms in the equation of motion of the scalar field we run a set of cosmological simulations using standard gravity and with the symmetron field in its static and non-static limits.  The static simulation was run using the solver that is described in detail in \citet[][]{2014A&A...562A..78L}.  The background evolution in all the simulations is given by a flat $\Lambda$CDM cosmology ($\Omega_m=0.267$, $\Omega_{\Lambda}=0.733$ and $H_0=71.9$ km/sec/Mpc).  The initial conditions were generated using Zeldovich approximation with standard gravity with the code mpgrafic \citep[][]{2008ApJS..178..179P}.  We use $512^3$ particles in a box of 128 Mpc/h.  The particular symmetron parameters employed are the same that were used by \citet[][]{2013PhRvL.110p1101L}: $z_{SSB}=0.5$ and $\lambda_0=1$ Mpc.  We also specified the coupling constant that was not employed by \citet[][]{2013PhRvL.110p1101L} to be $\beta=3$.  The nomenclature and details of the simulations is summarized in table \ref{table:simulations}.

\begin{table}[h!]
  \begin{center}
  \begin{tabular}{l|cc}
    name   & Solver & Refinements \\
    \hline 
    run\_newt & Newtonian & No \\
    run\_static & Symmetron static & No \\
    run\_full & Symmetron non-static & No \\
    \hline
    run\_newt\_ref & Newtonian & Yes \\
    run\_static\_ref & Symmetron static & Yes \\
  \end{tabular}
  \caption{Nomenclature and solvers used in the simulations.}
  \label{table:simulations}
  \end{center}
\end{table}

\citet[][]{2013PhRvL.110p1101L} have shown that the non-static solution for the symmetron scalar field contains patterns that do not strictly follow the density distribution (e.g.domain walls).  This is a situation that is highly difficult to handle with standard density based refinement criteria.  Therefore, we run the non-static simulation using only the domain grid of the code with $512^3$ nodes per dimension.  We determine the limitations of this approach, by comparing results from the Newtonian and static simulations with a pair of simulations that were run using seven levels of refinements (runs run\_newt\_ref and run\_static\_ref).

Domain walls are known to have their own dynamics, which can not be reproduced using a static solver.  In order to avoid biasing the results with unrealistic domain walls, we forced the static solver to provide a positive solution in the whole domain; in other words, we choose only one of the two possibles solutions when symmetry is broken.  On the other side, the non-static solution was left free to take negative and positive values.  

\section{Results}

\begin{figure}[t!]
\includegraphics[width=0.5\textwidth]{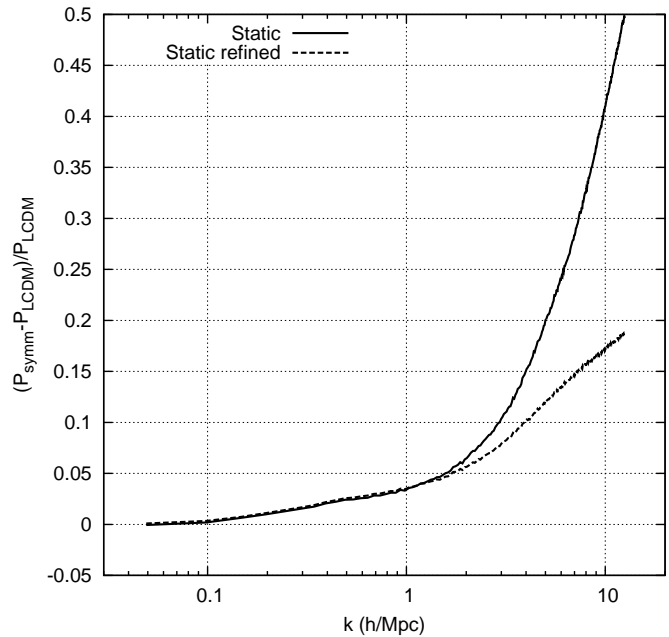}
\caption{Relative difference between the static symmetron and $\Lambda$CDM power spectra.}
\label{fig:power_static}
\end{figure}

\begin{figure*}
\begin{minipage}[t]{0.49\textwidth}
\includegraphics[width=1.0\textwidth]{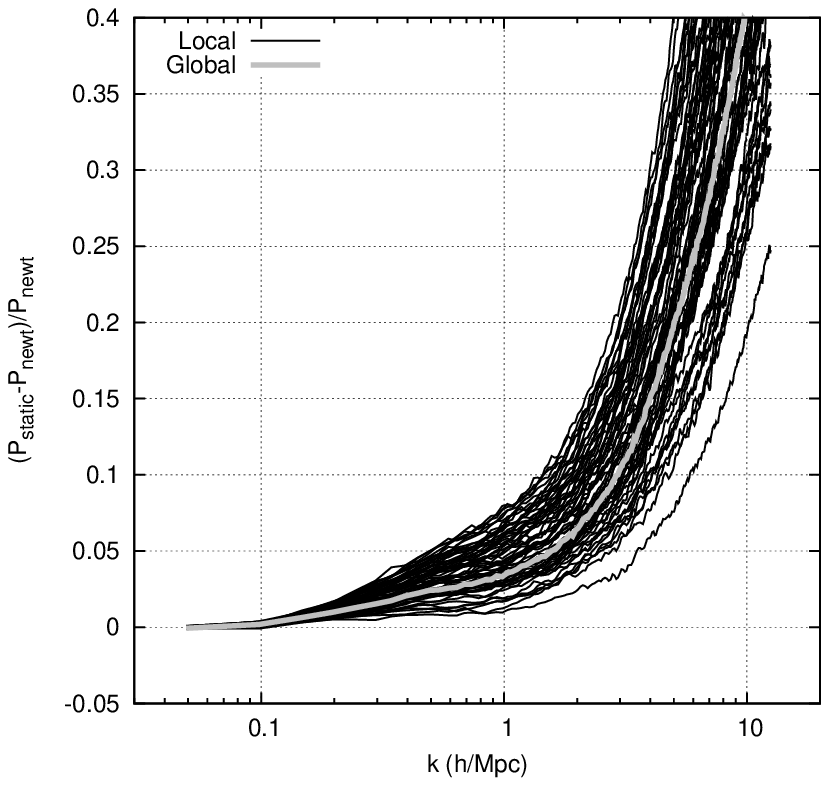}
\end{minipage}
\hfill{}
\begin{minipage}[t]{0.49\textwidth}
\includegraphics[width=1.0\textwidth]{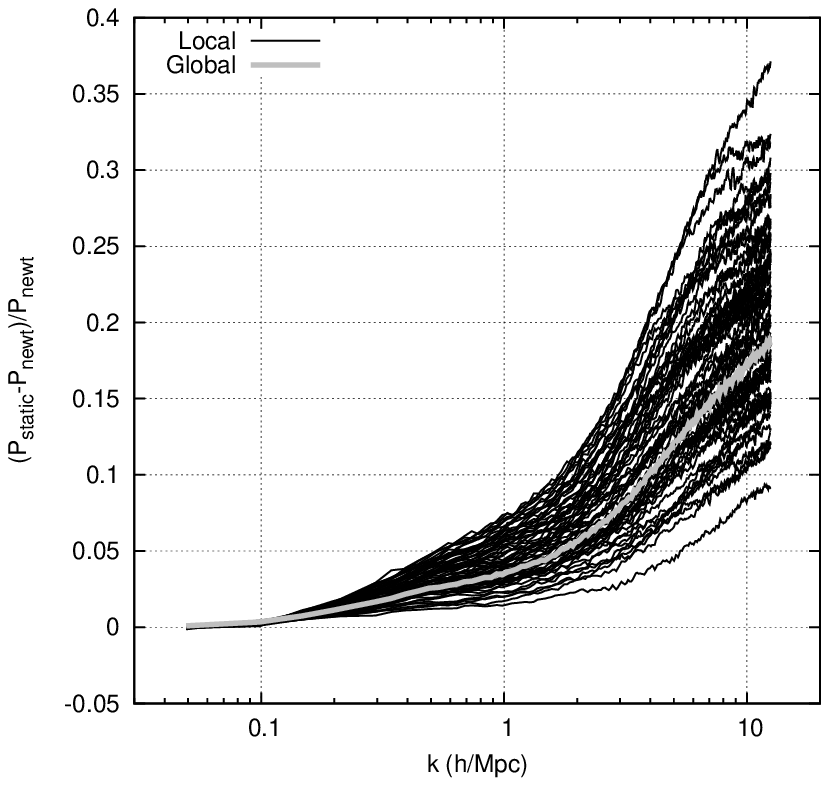}
\end{minipage}
\caption{Relative difference between power spectrum of static symmetron and $\Lambda$CDM simulations after filtering the density with gaussian distributions located in a grid of 4 nodes per dimension.  The thick gray line correspond to the global power spectrum.  The left and right panels correspond to non-refined and refined simulations respectively.}
\label{fig:power_static_gauss}
\end{figure*}

\begin{figure*}
\includegraphics[width=1.0\textwidth]{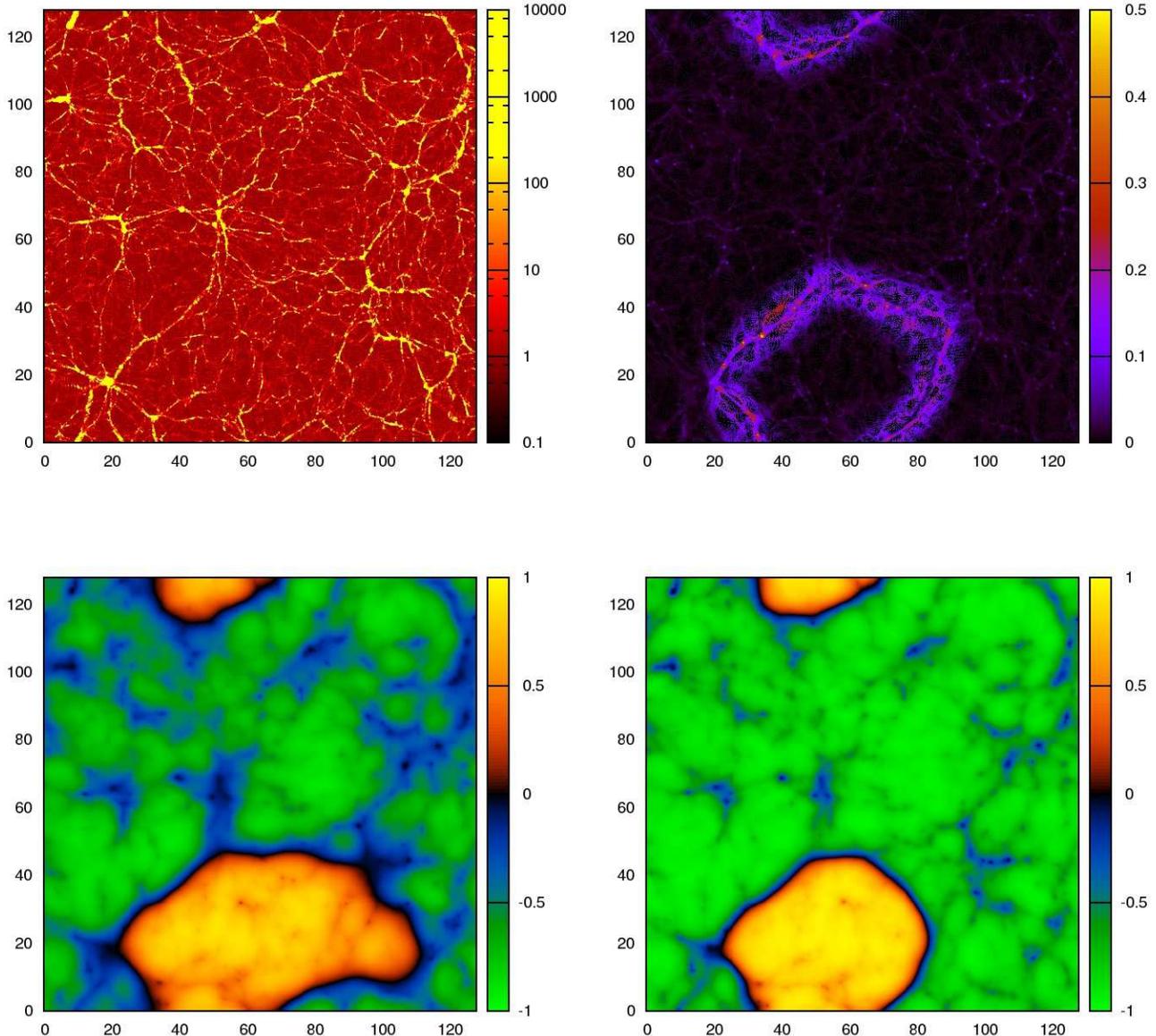}
\caption{Up-left: density distribution at redshift $z=0$ in a plane that passes though the center of the box (i.e. with coordinate $z=64$).  Up-right:  displacements between the position of the particles of the static and non-static runs at redshift $z=0$.  The particles shown belong to a slice of 2 Mpc/h thickness that passes through the center of the box.  Bottom:  scalar field at $z=0.58$ (left) and $z=0.32$ (right) in the same place shown in the upper panels.  See text for explanation.}
\label{fig:displacements}
\end{figure*}

\begin{figure}[t!]
\includegraphics[width=0.5\textwidth]{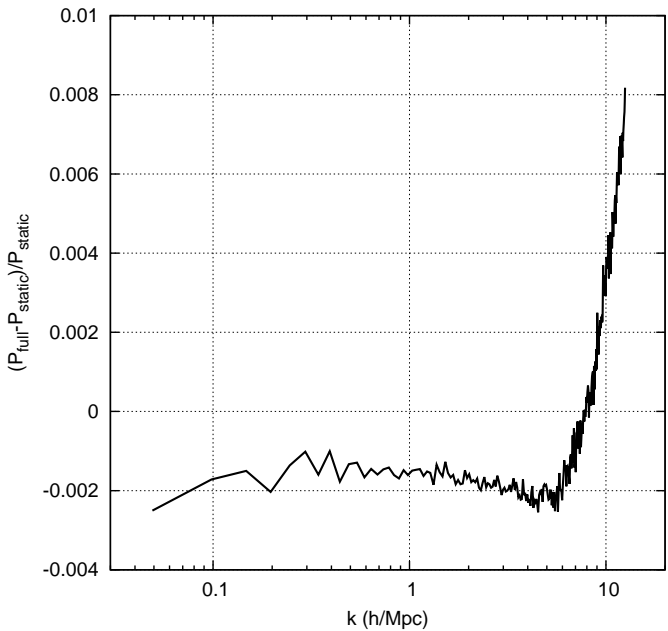}
\caption{Relative difference between non-static symmetron and Newtonian power spectra.}
\label{fig:power_full}
\end{figure}

\begin{figure}
\includegraphics[width=0.5\textwidth]{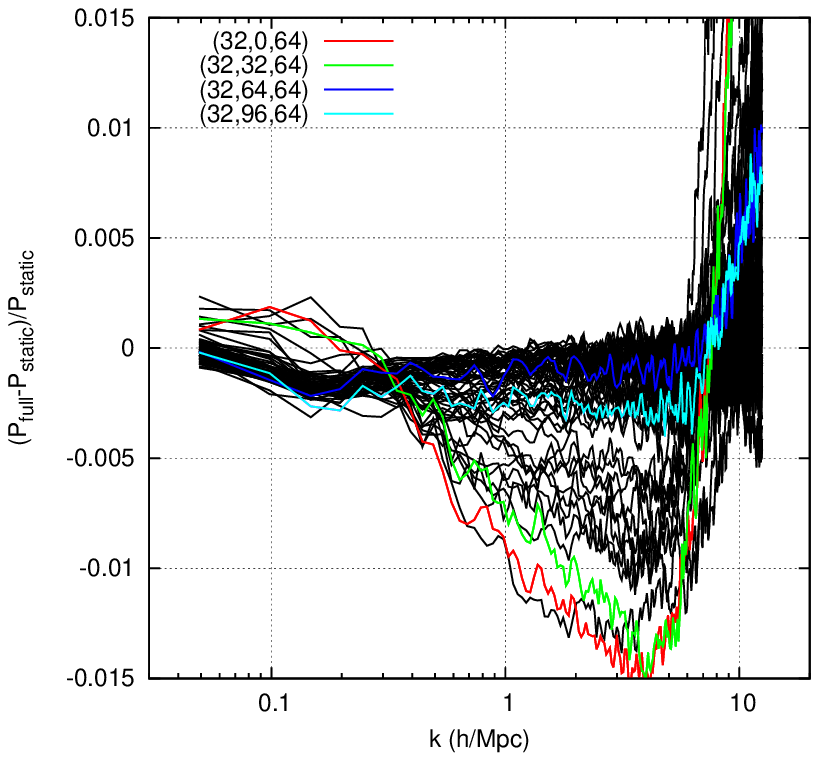}
\caption{Relative difference between power spectrum of static and non-static simulations after filtering the density with gaussian distributions located in a grid of 4 nodes per dimension.  The four color lines correspond to four representative locations which are inside and outside a domain wall.  Their positions of these particular filters are shown in units of Mpc/h in the upper-left region of the plot.  See upper-right panel of fig.\ref{fig:displacements} for comparison.}
\label{fig:power_full_gauss}
\end{figure}

\subsection{Static simulations}

Before to concentrate on the effects that the non-static terms of the Klein-Gordon equation have on the distribution of matter, we study differences between Newtonian and static symmetron evolution.  The impact of the non-static terms will be presented afterward as a correction to the differences found here.  We concentrate our study only in the global and local power spectrum.  The figure \ref{fig:power_static} shows the relative difference between the power spectrum of the static symmetron and $\Lambda$CDM simulations.  The estimation of the power spectrum was made using a grid with 512 nodes per dimension and following the Fourier based techniques and corrections presented in Ref. \citep[][]{2005ApJ...620..559J}.  The continuous line correspond to the difference between the simulations that do not include refinements (run\_newt and run\_static), while the dashed line is given by the refined simulations (run\_newt\_ref and run\_static\_ref).  The general behaviour of the power spectrum when including the fifth force is the same as shown for instance in Ref. \citep[][]{2012ApJ...748...61D}: there is an increase of the power on small scales, while the normalization given by the large scales is left unchanged.  The reason for the difference between the refined and non-refined simulation is that the resolution reached in the non-refined simulation is not enough high to resolve the screened region of the objects and thus the screening is less effective.  This gives higher values for the fifth force, which results in an increase in power at the smallest scales.  From the plot we can see that the non-refined simulations that will be used in the analysis of the non-static field can only be trusted up to frequencies $k\sim 2$.

The intensity of the effects associated to the symmetron model are know to have an environmental dependence (see for instance Ref. \citep{2012ApJ...756..166W}).  This means that the strength of the fifth force in a given halo is a function not only of its own matter distribution, but also of the surroundings in which the halo is immersed.  This bring us to a different observable that could be used to test the model, which is the localized power spectrum.  We calculated this quantity by filtering the density distribution before to calculate its Fourier transform by multiplying it with Gaussians centered at different positions in the box.  In other words, we calculated the power spectrum of the following overdensity distributions:
\be
\delta'(\textbf{x}) = (\delta(\textbf{x}) + 1) \times \exp(-(\textbf{x}-\textbf{x}_0)/(2\sigma^2)) - 1, 
\ee
where the overdensity $\delta$ is obtained from the position of the particles by means of a standard cloud in cell scheme \citep[e.g.][]{1988csup.book.....H}.  The dispersion that we choose is $\sigma=32$ Mpc/h, which corresponds to a quarter of the box size.  Fig.\ref{fig:power_static_gauss} shows the relative difference between this local power spectrum of the static symmetron and Newtonian simulations for 64 different positions $\textbf{x}_0$ of the gaussian filter.  The values of $\textbf{x}_0$ are given by a 3D uniform grid of four nodes per dimension.  The gray thick line corresponds to the global power spectrum already shown in fig.\ref{fig:power_static} while the left and right panels are results from the non-refined and refined simulations respectively.  Owing to the fact that we are studying the relative difference with $\Lambda$CDM simulations and that the initial conditions are exactly the same for all the simulations, the plots should not be affected by cosmic variance.  The dispersion that can be seen between the power spectra at different positions is physical and related to the environmental dependence of the fifth force.

\subsection{Effects of non-static terms in the distribution of matter}

We now concentrate on the effects that the time derivatives of the scalar field and the presence of domain walls have in the matter distribution at redshift $z=0$.  The Fig. \ref{fig:displacements} shows the impact of the domain walls formed in the simulation run\_full on the matter distribution.  The upper-left and bottom panels show the over-density (top) and scalar field (bottom) in plane that passes through the center of the box extracted from the grid that was used during the simulation.  The scalar field $\chi$ is shown at two different redshifts (z=0.583 to the left and z=0.315 to the right), while the density distribution is given at redshift $z=0$.  It is possible to see the formation of a wall which survives from a redshift close to $z_{SSB}$ (when the scalar field starts to oscillate away from zero) up to $z\sim 0.3$.  After that moment, the wall collapses releasing its energy in scalar waves.  The plot shows that the spatial configuration of the wall is not independent of matter as standard domain walls, but that the coupling to matter in the Klein-Gordon equation makes it to be more stable in places where the symmetry is restored.  Thus, the domain walls follow closely the distribution of halos and filaments.  

The effects that the fifth force associated with this domain wall has on the matter distribution can be seen in the top-right panel of the same figure.  In there, we plot color coded the absolute displacement between the particles of the static and non-static symmetron simulations (run\_static and run\_full) at the position of each particle in a 2 Mpc/h thick slice that passes through the center of the box.  It is possible to see that the larger displacements (up to half a Mpc/h) occur in the position of the wall.  While the wall does not survive until redshift zero, the extra kick that it gives to the particles during its relatively short life time is enough to produce changes in the density distribution that can last until redshift $z=0$.  Note that the fifth force in the symmetron model is proportional to $\nabla\chi^2$ (as opposed to $\nabla\chi$) and thus, the domain walls do not push in only one direction, but have an attractive behaviour.

In order to check the impact that these differences found in the position of the particles have in statistical observables, we calculated and compare the global and local power spectra in the same way as done before for the static simulation.  Here, instead of comparing against the Newtonian simulation, we do the comparison between the static and non-static simulations, without making reference to the Newtonian one.  The figure \ref{fig:power_full} shows the relative difference between the power spectrum of these two simulations.  The differences are below the percent level in the whole domain of scales that we study.  The apparent offset of 0.2 \% has contributions from two separate effects.  The small scale offset is physical produced by the presence of the domain wall.  On the other hand, the large scale offset is numerical.  It is related to the fact that the routine \texttt{advance\_chi} (see pseudo-code) uses values of the density that were obtained at the begining of the large time step, while the static code uses the old and new density in each half of the time steps.  In any case, the differences are negligible in the sense that are beyond the precision that can be reached with present and near future observations.  Thus, the static simulation technique is safe in case that only the global power spectrum is to be calculated (at least within the range of model parameters that are close to the once studied here).  Extension of this result to more general models is left for future work.

  The situation is different when local perturbations are studied.  Fig.\ref{fig:power_full_gauss} shows the relative difference between the local power spectrum of the static and non-static simulations calculated in the same way as done for the comparison between static and Newtonian simulations.  Now it is possible to see differences of the order of 1\% in the region in which the simulations can be trusted (i.e. the region where refined and non-refined simulations give consistent power spectra).   To clarify the reason for the differences found, we highlighted with color lines the curves corresponding to representative places that lie inside and outside a domain wall (see figure \ref{fig:displacements} for reference).  The presence of the domain wall is responsible for the lack of power in the non-static simulations.

\section{Conclusions}

This work is a companion paper to our earlier article \citep[][]{2013PhRvL.110p1101L} where we presented a new N-body code within the framework of
scalar-tensor theories which takes into account
the temporal derivaties of the scalar degree of freedom. 
Previous works on N-body cosmological simulations with scalar-tensor theories for gravity have the static approximation as their main assumption, which means that the time derivatives in the Klein-Gordon equation are neglected.  
%The approximation is known to be valid for standard gravity.  However, the impact of non-static effects in the modified gravity case is still unclear, especially in the non-linear regime of cosmological evolution. 
The impact of non-static effects is still unclear, especially in the non-linear regime of cosmological evolution.  Here we propose to test the validity of this assumption by running cosmological simulations including non-static terms in the Klein-Gordon equation for the scalar field.  Our analysis is based in the symmetron model, however our techniques can be easily generalized to others.

A large part of this paper is devoted to the description of the algorithms that we use to solve the complete Klein-Gordon equation.  The paper also presents the modifications that we made to the Isis-Ramses N-body code as well as the tests that we made to confirm that these modifications were properly implemented.

We determine the importance of non-static effects by comparing results obtained with this new code with static simulations that were run using the static solver presented in Ref. \citep[][]{2014A&A...562A..78L}.  Non-static cosmological simulations were already reported by \citet[][]{2013PhRvL.110p1101L}.  However, that particular study concentrated only in the properties of the non-static solutions for the scalar field and there was no mention to the impact that these new solutions have on the matter distribution.  Here we run similar simulations, but including the fifth force in the geodesics equation.  Furthermore, we increased the resolution by a factor of four with respect to this previous study.  

In first place, we studied the impact that the static fifth force has on the matter distribution with respect to standard Newtonian gravity.  To this end, we used the static symmetron code that is described in detail in \citet[][]{2014A&A...562A..78L}.  We studied the global and local power spectrum and found that, for the particular set of parameters used in this paper, there is a global increase of the power which can go up to 20\% at the smallest scales studied.  The local power spectrum shows that environmental effects gives a large variance to the power spectrum in addition to the known cosmic variance.  In other words, the extra bit of evolution produced by the symmetron field is a function of the position in space.

The non-static simulation that we run with the new non-static code recovers results of previous study in the sense that the scalar field develops domain walls which not only have a dynamics that can not be recovered with static solutions, but that can also collapse releasing their energy in scalar waves.  Regarding the effects of non-static terms in the matter distribution, we found almost no deviation in the global matter power spectrum between the static and non-static simulations.  However, we find that the fifth force induced by the domain walls do change the distribution of matter.  This effects can be seen in the local power spectrum, which shows deviations of the order of 1\% in the region of the frequency space in which we can trust the simulations.  For comparison, note that for this frequencies, the local power spectrum have a dispersion of around 5\% when comparing the symmetron and Newtonian simulations in the static limit.

Given the results that we obtained for the symmetron model, we can attempt to extrapolate our conclusions to similar scalar-tensor models such as chameleons or galileons.  Our analysis of the local power spectrum shows that the larger differences between static and non-static power spectra are produced by the presence of domain walls, which do not form in any of these other models.  The global power spectrum shows that the domain walls affect the spectrum in only 0.2 per cent.  Thus, we expect that usual cosmological probes such as power spectrum for present day experiments will not have enough precision to detect any difference.  However, there may be other observables which can be affected by the non-static terms.

\section{Acknowledgments}

We thank the Research Council of Norway FRINAT grant 197251/V30.  CLL acknowledges hospitality at Universitat de Barcelona.  The simulations were performed on the NOTUR Clusters \texttt{HEXAGON}, the computing facilities at the Universities of Bergen, Norway.  We thank Iain Brown for helpful discussions.

\bibliography{references}

\end{document}